\documentclass[twocolumn]{aastex62}
\submitjournal{ApJ}

\usepackage{amsmath,natbib, float, color, soul, xspace}
\usepackage[normalem]{ulem}

\newcommand*\Msolarh[0]{h^{-1} \, \mathrm{M_{\odot}}}

\newcommand*\chandra{\textit{Chandra}\xspace}
\newcommand*\tng{\textit{IllustrisTNG}\xspace}

\newcommand*\new[1]{#1}

\newcommand{\code}[1]{\texttt{#1}}

\shorttitle{Cluster X-ray Masses with ML}
\shortauthors{Ntampaka, et al.}

\begin{document}

\correspondingauthor{Michelle Ntampaka}
\email{michelle.ntampaka@cfa.harvard.edu}

\author{M. Ntampaka}
\affiliation{{Center for Astrophysics $|$ Harvard \& Smithsonian, Cambridge, MA 02138, USA}}
 \affiliation{Harvard Data Science Initiative,
  Harvard University,
  Cambridge, MA 02138, USA}

\author{J. ZuHone}
\affiliation{Smithsonian Astrophysical Observatory,
 Cambridge, MA 02138, USA}
 
\author{D. Eisenstein}
\affiliation{{Center for Astrophysics $|$ Harvard \& Smithsonian, Cambridge, MA 02138, USA}}
 
\author{D. Nagai}
\affiliation{Department of Physics, Yale University, New Haven, CT 06520, USA}

\author{A. Vikhlinin}
\affiliation{{Center for Astrophysics $|$ Harvard \& Smithsonian, Cambridge, MA 02138, USA}}
\affiliation{Space Research Institute (IKI), Profsoyuznaya 84/32, Moscow, Russia}

\author{L. Hernquist}
\affiliation{{Center for Astrophysics $|$ Harvard \& Smithsonian, Cambridge, MA 02138, USA}}

\author{F. Marinacci}
\affiliation{{Center for Astrophysics $|$ Harvard \& Smithsonian, Cambridge, MA 02138, USA}}

\author{D. Nelson}
\affiliation{Max-Planck-Institut f{\"u}r Astrophysik, Karl-Schwarzschild-Stra{\ss}e 1, D-85741, Garching bei M{\"u}nchen, Germany}

\author{R. Pakmor}
\affiliation{Max-Planck-Institut f{\"u}r Astrophysik, Karl-Schwarzschild-Stra{\ss}e 1, D-85741, Garching bei M{\"u}nchen, Germany}

\author{A. Pillepich}
\affiliation{Max-Planck-Institut f{\"u}r Astronomie, K{\"o}nigstuhl 17, D-69117, Heidelberg, Germany}

\author{P. Torrey}
\affiliation{Department of Astronomy, University of Florida, 211 Bryant Space Sciences Center, Gainesville, FL 32611, USA}

\author{M. Vogelsberger}
\affiliation{Kavli Institute for Astrophysics and Space Research, Massachusetts Institute of Technology, Cambridge, MA 02139, USA}

\title{A Deep Learning Approach to Galaxy Cluster X-ray Masses}

\begin{abstract}
We present a machine-learning approach for estimating galaxy cluster masses from \chandra mock images.  We utilize a Convolutional Neural Network (CNN), a deep machine learning tool commonly used in image recognition tasks.  The CNN is trained and tested on our sample of 7,896 \chandra X-ray mock observations, which are based on 329 massive clusters from the \tng simulation.  
Our CNN learns from a low resolution spatial distribution of photon counts and does not use spectral information.  Despite our simplifying assumption to neglect spectral information, the resulting mass values estimated by the CNN exhibit small bias in comparison to the true masses of the simulated clusters (-0.02 dex) and reproduce the cluster masses with low intrinsic scatter, 8\% in our best fold and 12\% averaging over all.  In contrast, a more standard core-excised luminosity method achieves 15-18\% scatter. We interpret the results with an approach inspired by Google DeepDream and find that the CNN ignores the central regions of clusters, which are known to have high scatter with mass.   \\ 
\end{abstract}

%text abstract: We present a machine-learning approach for estimating galaxy cluster masses from \chandra mock images.  We utilize a Convolutional Neural Network (CNN), a deep machine learning tool commonly used in image recognition tasks.  The CNN is trained and tested on our sample of 7,896 \chandra X-ray mock observations, which are based on 329 massive clusters from the \tng simulation.  Our CNN learns from a low resolution spatial distribution of photon counts and does not use spectral information.  Despite our simplifying assumption to neglect spectral information, the resulting mass values estimated by the CNN exhibit small bias in comparison to the true masses of the simulated clusters (-0.02 dex) and reproduce the cluster masses with low intrinsic scatter, 8\% in our best fold and 12\% averaging over all.  In contrast, a more standard core-excised luminosity method achieves 15-18\% scatter. We interpret the results with an approach inspired by Google DeepDream and find that the CNN ignores the central regions of clusters, which are known to have high scatter with mass.   


\section{Introduction}
\label{sec:intro}

Galaxy clusters are gravitationally bound systems that contain hundreds or thousands of galaxies in dark matter halos of mass $\gtrsim 10^{14}\, M_\odot$.  They are massive and rare, and their abundance is sensitive to the underlying cosmological model.  Utilizing cluster abundance as a cosmological probe requires a large cluster sample with a well-defined selection function, a way to connect observations to the underlying dark matter, and an understanding of the scatter in the mass-observable relationship.

A number of mass proxies can be derived from X-ray observations of galaxy clusters.  X-ray cluster observations probe a portion of the baryonic component of clusters \textemdash{}  the hot intracluster medium \textemdash{}  which emits X-ray radiation primarily through bremsstrahlung.  Hydrodynamical simulations \citep{2014ApJ...782..107N, 2016MNRAS.455.1115H, 2017MNRAS.466.4442L, 2018MNRAS.477.3727B, 2018MNRAS.476.2999M} can be used to model observable-mass relations in clusters.

Cluster luminosity ($L_X$) is correlated with mass, and excising the inner $\approx 0.15\,R_{500c}$ reduces scatter further due to variations in the core properties \citep{2007ApJ...668..772M, 2018MNRAS.473.3072M}.  The global temperature ($kT$) relates to cluster mass through the virial theorem, scaling with mass as a power law \citep[e.g.~][]{2005A&A...441..893A}.  {For long exposures of bright, low-redshift clusters, it is possible to access luminosity and spectral cluster profiles, leading to tighter mass-observable relationships.  Hydrostatic mass estimates can be calculated from temperature and density gradients \citep{2005ApJ...628..655V} and the product of spectral temperature ($T_X$) and gas mass ($M_g$)  denoted $Y_X$, is a very low-scatter mass proxy, with intrinsic scatter of $\approx5\%-7\%$ \citep{2006ApJ...650..128K}.  
However, the hydrostatic mass estimates are known to be biased \citep[e.g.][]{2007ApJ...655...98N}, and it is one of the primary sources of systematic uncertainties in the cluster-based cosmological measurements \citep{2016A&A...594A..24P}. }

{  A number of other physical processes impact X-ray based cluster mass estimates, including non-thermal pressure \citep{2009ApJ...705.1129L, 2014ApJ...792...25N}, gas clumping \citep{2011ApJ...731L..10N}, temperature inhomogeneities \citep{2014ApJ...791...96R}, and cluster dynamical state \citep{2008ApJ...685..118V, 2012ApJ...754..119M}.
Calculable morphological parameters correlate with dynamical state, including surface brightness concentration \citep[e.g.][]{2008A&A...483...35S}, centroid shift \citep[e.g.][]{2016MNRAS.457.4515R}, and morphological composite parameters \citep[e.g.][]{2013AstRv...8a..40R}.  These suggest that cluster observables are tied to mass in a complex way that may be exploited to reduce scatter and improve individual cluster mass estimates.}

In addition to affecting mass estimate errors, a cluster's dynamical state influences the probability that the cluster will be observed.  Sunyaev-Zeldovich \citep[SZ,][]{1972CoASP...4..173S}-selected samples preferentially have more disturbed clusters \citep{2011A&A...536A...9P}, while X-ray-selected samples have a higher fraction of relaxed clusters \citep{2011A&A...526A..79E}.  If the fractions of relaxed and disturbed systems in cluster samples are not well known, this may introduce a bias \citep{2002ApJ...577..579R}.  

\begin{figure*}[]
	\begin{center}
		\includegraphics[width=0.9\textwidth]{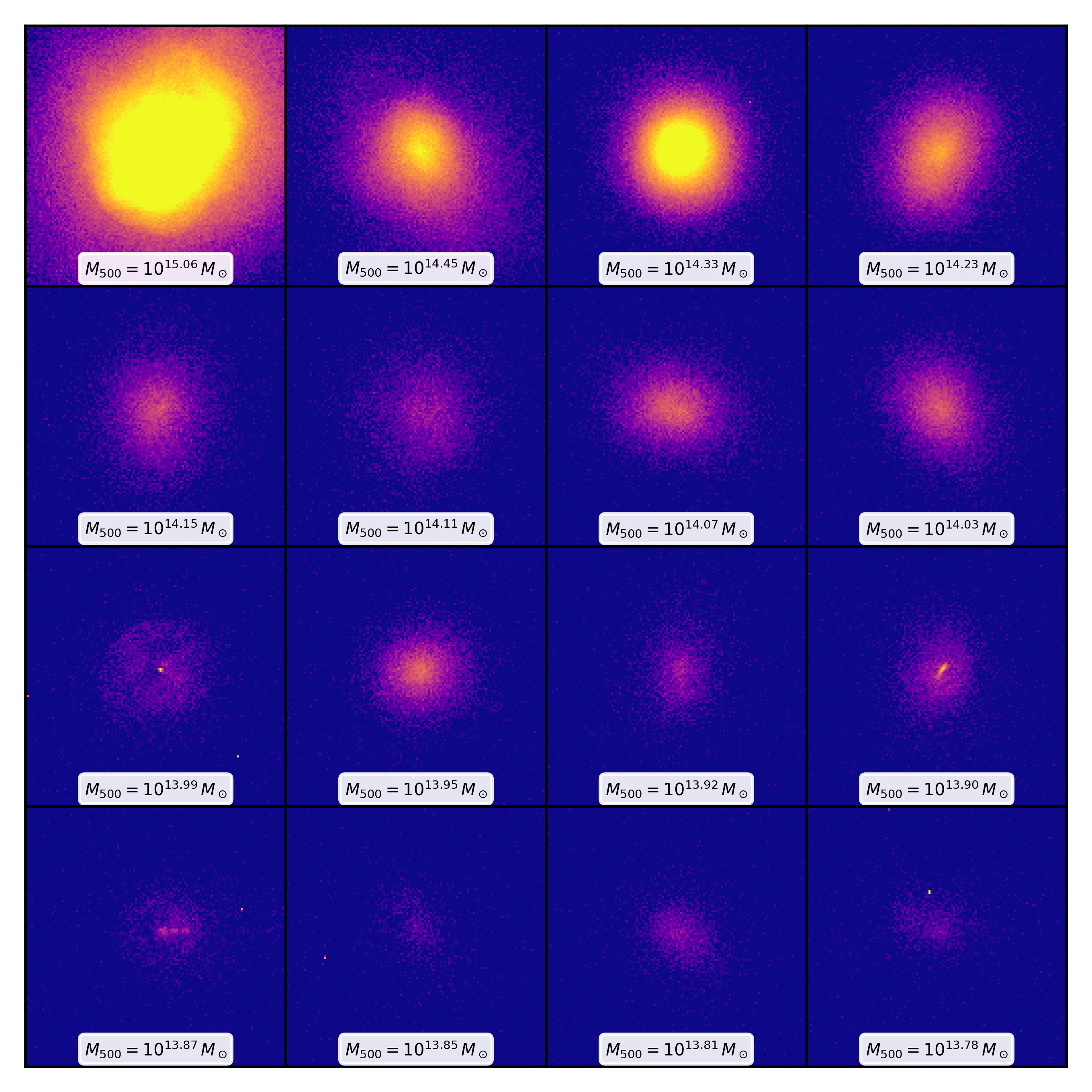} \\
		\includegraphics[width=0.935\textwidth]{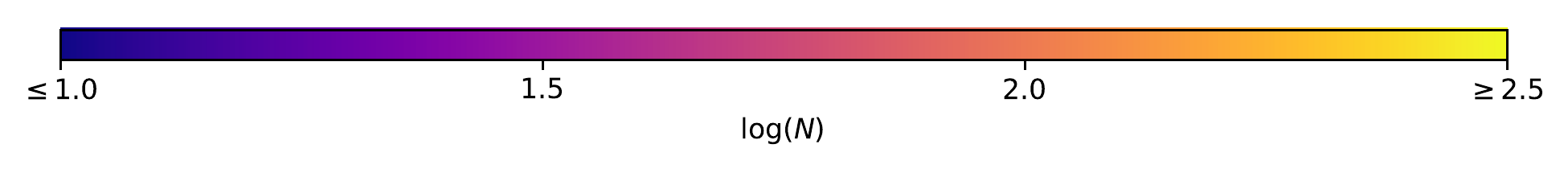} \\
	\caption[]{A sample of 16 of the 7,896  mock X-ray cluster observations created with pyXSIM software applied to the \tng cosmological hydrodynamical simulation.  The mock observations emulate 100ks \chandra observations that have been degraded to $128\times128$ postage stamp images for one broad ($0.5-7\mathrm{keV}$) energy band; shown are the number of photons, $N$, in each pixel for this mock observation.  Each unique cluster in the simulation is used to produce 24 mock images according to the data augmentation scheme described in Section \ref{sec:chandra}.   }
       	\label{fig:sampledat}
      	\end{center}
\end{figure*}

Machine learning (ML) offers a number of tools that can be used to untangle subtle signals and extract complicated correlations.   ML has been utilized in astronomy and cosmology for classification tasks such as 
labeling galaxy morphology \citep{2010MNRAS.406..342B, 2015MNRAS.450.1441D, 2018MNRAS.476.3661D}, identifying transient types \citep{2015AJ....150...82G}, identifying the presence or absence of lensing signals in images \citep{2018MNRAS.473.3895L}, categorizing the type of sources driving reionization \citep{2018arXiv180703317H}, and estimating photometric redshifts \citep{2018arXiv180606607P}.  ML has also been used for astronomical and cosmological regression tasks, for example, reducing errors in cluster dynamical mass measurements \citep{2015ApJ...803...50N, 2016ApJ...831..135N, 2018arXiv181008430A}, {determining the duration of reionization \citep{2018arXiv181008211L}}, and producing tighter cosmological parameter constraints with mock catalogs \citep{2018PhRvD..97j3515G}.

Because ML has been successful in harnessing complicated correlations in other astronomical applications, they may be useful in using subtle signals in X-ray images to improve mass estimates.  One class of ML algorithms that has had much success in image-based tasks are Convolutional Neural Networks \citep[CNNs, e.g.~][]{fukushima1982neocognitron, lecun1999object, NIPS2012_4824, DBLP:journals/corr/SimonyanZ14a}.  CNNs have many hidden layers, often pairing layers of convolution and pooling to extract features from the input images.  They require very little preprocessing of the input images because the network learns the convolutional filters necessary to extract relevant features.  See \cite{2014arXiv1404.7828S} for a review of deep neural networks.

{We present a method for predicting cluster masses from decreased-resolution \chandra mock X-ray images that utilizes a CNN.}  We describe the mock  observations in Section \ref{sec:chandra} and the CNN method and architecture in Section \ref{sec:cnn}.  We show the resulting mass predictions in Section \ref{sec:results}, interpret the model in Section \ref{sec:interpret}, and conclude in Section \ref{sec:conclusion}.

\section{Methods}
\label{sec:methods}

\subsection{Mock \chandra Observations}
\label{sec:chandra}

\subsubsection{IllustrisTNG Clusters}
\label{sec:tng}

A sample of  simulated clusters is drawn from the \tng  cosmological hydrodynamical simulation \citep{2018MNRAS.475..624N, 2018MNRAS.475..676S, 2018MNRAS.480.5113M, 2018MNRAS.477.1206N, 2018MNRAS.475..648P}.  \tng uses an updated galaxy formation model \citep{2017MNRAS.465.3291W, 2018MNRAS.473.4077P} to overcome many of the physical limitations of the previous \textit{Illustris} model \citep{2013MNRAS.436.3031V, 2014MNRAS.438.1985T, 2014MNRAS.445..175G, 2014MNRAS.444.1518V, 2014Natur.509..177V, 2015A&C....13...12N}.  The suite of \textit{IllustrisTNG} simulations assume a $\Lambda$CDM cosmology with parameters consistent with \cite{2016A&A...594A..13P}. With a simulated cubic volume of $300\,\mathrm{Mpc}$ on a side, \textit{TNG300} is the largest of the suite, making it ideal for studying rare and massive clusters.  Furthermore, the simulation is performed at an unprecedented resolution, with baryonic mass resolution of $7.6\times10^{6}\, M_\odot$ \citep{2018MNRAS.475..624N}.

We select 329 massive clusters within a mass range of $M_{500c}=10^{13.57}$ to $M_{500c}=10^{15.06}$ from the \textit{TNG300} simulation, using the Friends-of-Friends (FoF) ``group'' \citep{1985ApJ...292..371D} halos from the $z = 0$ snapshot.  While an arbitrary spherical overdensity halo definition may include particles not linked within this FoF group, we are interested in predicting $M_{500c}$ and the associated $R_{500c}$ is small enough that all gas within this radius also should be found in the FoF group.  Every gas cell associated with each group is included, so that all of the substructures associated with each cluster are used in the computation of the X-ray emission.

\subsubsection{pyXSIM}
\label{sec:pyxsim}

Our mock X-ray observations of the \tng cluster sample are produced using the \code{pyXSIM}\footnote{\url{http://hea-www.cfa.harvard.edu/~jzuhone/pyxsim/}} \citep{zuhone-proc-scipy-2014} and \code{SOXS}\footnote{\url{http://hea-www.cfa.harvard.edu/~jzuhone/soxs/}} software packages. {\code{pyXSIM} is an implementation of the \code{PHOX} algorithm \citep{2013MNRAS.428.1395B, 2012MNRAS.420.3545B}.}  Large photon samples are initially built in \code{pyXSIM} from the 3D distributions of density, temperature, and metallicity from the \tng data for each cluster using an APEC emission model \citep{2012ApJ...756..128F}, assuming a redshift of $z = 0.05$. Only particles with $kT > 0.1$~keV {that are not forming stars are} used in the construction of the photon samples. {These samples are then projected along each of the $x$-, $y$-, and $z$- axes of the simulation box}, and foreground galactic absorption is applied to each sample assuming the \code{wabs} \citep{1983ApJ...270..119M} model with a value of $N_H = 4 \times 10^{20} cm^{-2}$ for each cluster. 

\begin{figure*}[]
	\begin{center}
		\includegraphics[width=\textwidth]{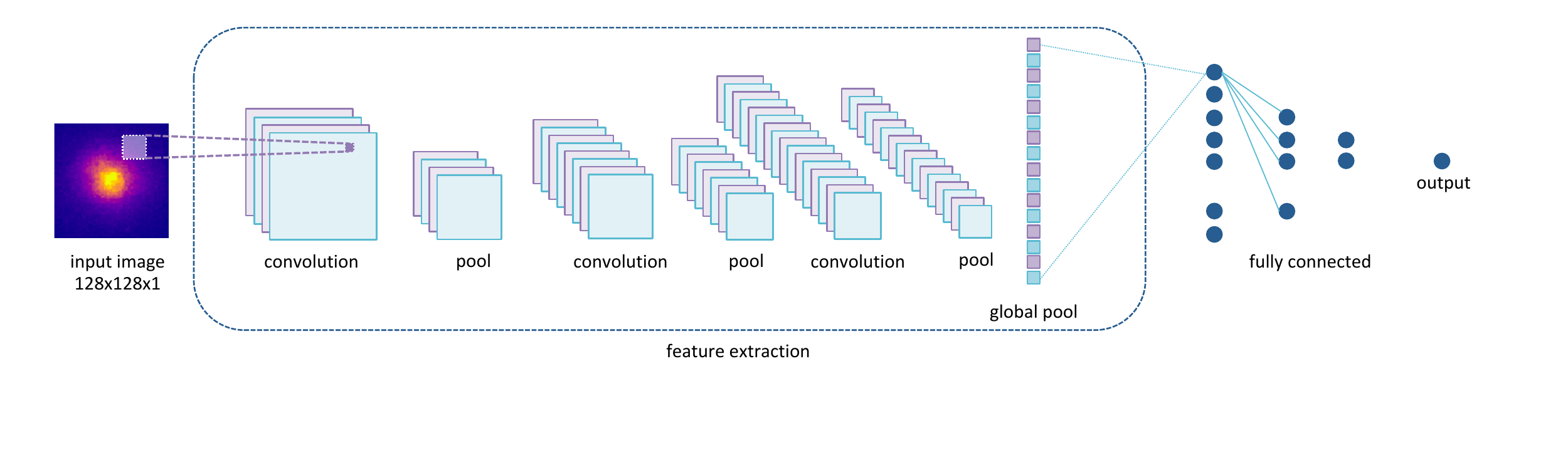} \\
	\caption[]{Architecture of the Convolutional Neural Network (CNN) used in this analysis.  Our network utilizes three convolutional and pooling layers for feature extraction and three fully connected layers for parameter estimation.}
       	\label{fig:architecture}
      	\end{center}
\end{figure*}

Each photon sample is then convolved with an instrument model for \chandra's ACIS-I detector using the \code{SOXS} package. We assume a simplified representation of the ACIS-I detector, with a 20' square field of view without chip gaps and 0.5" pixels. {At our cluster sample redshift, $R_{500c}$ extends beyond the 20' square field of view of the detector for clusters with mass $M_{500c}\gtrsim10^{13.8}\,M_\odot$.  Rather than producing a tiled \chandra image that fully contains $R_{500c}$, we opt to use mock cluster observations that could be achieved from a single pointing.} The PSF is Gaussian-shaped with FWHM 0.5", and the effective area and spectral response are taken from the Cycle 19 aimpoint response files (ARF and RMF) and assumed to be the same across the entire detector. The built-in models for the ACIS-I particle background and the galactic foreground included with \code{SOXS} were also applied\footnote{\url{http://hea-www.cfa.harvard.edu/~jzuhone/soxs/users_guide/background.html}}. We integrate each observation for 100~ks.

\newpage

\subsubsection{Image Preprocessing}
\label{sec:preprocess}

\new{Because the machine learning tool described in Section \ref{sec:cnn} is not invariant under image rotation}\footnote{\new{Rotationally invariant CNNs are an area of active research, see, for example, \cite{2015MNRAS.450.1441D} and \cite{worrall2017harmonic}}}, we augment the data set with $90^\circ$ rotations as well as reflections along the vertical and horizontal axes \new{\citep[as in, e.g.,][]{2017ApJ...836...97C}}.  The three 2D projections, two axial reflections, and four possible $90^\circ$ orientations result in 24 images for each unique cluster.  All 24 images of each unique cluster are assigned to one of 10 groups, called folds.  The clusters are ordered by mass and cyclically assigned to folds so that each fold has approximately the same mass function.

Each mock \chandra event file is degraded in spatial resolution to a $128\times128$ ``postage stamp'' image over the broad energy band of 0.5-7~keV. {The decreased resolution has two advantages:  it decreases computation time and it also decreases the effects of the nonuniform distortion of the \chandra PSF.  Moving to a lower resolution reduces the effects of the nonuniform PSF in a real observation, making it unnecessary to model it precisely.} {Example images of a representative sample of 16 clusters spanning the mass range are shown in Figure \ref{fig:sampledat}.} The final catalog is comprised of 24 decreased-resolution \chandra images of each of 329 unique \tng clusters, totaling 7,896 \chandra mock observations.

\subsection{Convolutional Neural Networks}
\label{sec:cnn}

Convolutional Neural Networks \citep[CNNs, ][]{fukushima1982neocognitron, lecun1999object, NIPS2012_4824} are a class of feed-forward machine learning algorithms that are commonly used in image recognition tasks.  They use pairs of convolutional filters and pooling layers to extract meaningful patterns from the input image, and {can be used for both classification and regression tasks.}  Because the network learns the convolutional filters, CNNs require very little preprocessing of the input images.

{The CNN is implemented in Keras \citep{chollet2015} with a Tensorflow \citep{45381} backend.}  \new{Our CNN architecture is based loosely on a simplified version of \cite{DBLP:journals/corr/SimonyanZ14a} with fewer hidden layers}; it is shown in Figure \ref{fig:architecture}.  {The model is implemented with sequential layers as follows:
	\begin{enumerate}
	\itemsep-0.5em  
		\item $3\times 3$ convolution with 16  filters
		\item $2\times 2$, stride-2 max pooling 
		\item $3\times 3$ convolution with 32  filters
		\item $2\times 2$, stride-2 max pooling 
		\item $3\times 3$ convolution with 64  filters
		\item $2\times 2$, stride-2 max pooling 
		\item global average pooling
		\item 10\% dropout
		\item 200 neurons, fully connected 
		\item 10\% dropout
		\item 100 neurons, fully connected 
		\item 20 neurons, fully connected 
		\item output neuron
	\end{enumerate}
We use a  mean squared loss function and the Adam Optimizer \citep{2014arXiv1412.6980K} with learning rate reduced to half of the default value ($\mathrm{lr}=0.0005$).  } 

{In our model,} feature extraction is performed by three pairs of $3\times3$ convolutional filters coupled with $2\times2$ stride-2 max pooling layers \citep[e.g.][]{riesenhuber1999hierarchical}.  These are followed by a global average pooling layer \citep{2013arXiv1312.4400L} and three fully connected layers with rectified linear unit \citep[ReLU,][]{nair2010rectified} activation.  \new{A 10\% dropout after two fully connected layers prevents overfitting \citep{srivastava2014dropout}.}  

\new{For the task of regressing a single parameter, we use an architecture with one output neuron.  This output neuron gives a continuous-valued label (regression) rather than a class probability (classification).  We select the  mean squared loss function for this regression task.  Our model has 58,437 tunable parameters; because of the global average pooling layer that compresses the information into 64 neurons, the number of tunable parameters is invariant to changes to the input image resolution. }

\new{We use the $128\times128$ images as input and train the model to predict $\log(M_{500c})$ from these images.} \new{We perform a 10-fold crossvalidation, dividing the sample into a training set that comprises 80\% of the images, a validation set that comprises 10\% of the images, and a test set of the remaining 10\% of the images.}  Every rotation, axial flip, and line of sight view of a single cluster is assigned to only one of these sets; a cluster is never used, for example, to train a model and subsequently test it.
\new{The training set is used to train the model to minimize a mean squared error loss function, the validation set is used to assess the stopping criteria, and the mass predictions for the test set are reported.}  We cycle through the data assignments to the train, validation, and test folds until the masses of all clusters have been predicted. 

Stopping criteria are implemented to converge on models that, when applied to the validation set, are low scatter, low bias, and have no catastrophic outliers.  The mass residual, $\delta$, is defined as
\begin{equation}
	\delta \equiv \log(M_\mathrm{predicted}) - \log(M_\mathrm{true}).
\end{equation}
We consider the training of the CNN to have converged when the following three criteria are met:  
maximum absolute value of mass residual  is less than 0.3,
absolute value of median residual error is less than 0.02, 
slope of the best fit line of $M_\mathrm{true}$ vs. $M_\mathrm{predicted}$ is greater than 0.9. 
 For several folds, these three criteria were not met, so the model at the 400th training cycle, or ``epoch,'' is used.  In most cases, the training converged within 150 epochs.  
We emphasize that these criteria are applied to the validation set and these criteria may not be descriptive of the test set.

\section{{Results}}
\label{sec:results}

The mass predictions for all 10 folds are shown in Figure \ref{fig:scatter}.  Vertical streams of points show the mass estimates for each of the 24 images of each cluster.  
The results show a tendency to predict toward the mean, evident in the overprediction of low mass clusters and the underprediction of high mass clusters.  When applying this method to a sample of observed clusters, one would create a training catalog that extends beyond the estimated mass range of the test catalog to mitigate this issue.

A PDF of mass errors is shown in Figure \ref{fig:errpdf}.  The 10-fold distribution is fit to a Gaussian with width $\sigma=0.051$ dex (corresponding to a $11.6\%$ scatter) and a small negative bias given by the mean $\mu=-0.022$ dex.   In practice, the train set should extend well beyond the estimated mass range of test clusters, using the model to make predictions for clusters at the middle mass region to mitigate the effects of bias to the mean.  In the mass range $14.0<\log(M_{500c})<14.5$, the 10-fold distribution is well-described by a Gaussian with width $\sigma=0.41$ dex, $9.6\%$ scatter, and bias $\mu=-0.036$ dex.  

The initial random state of a CNN can affect the solution upon which it converges, and so it is informative not only to evaluate the 10-fold mean, but also to evaluate the best-fit fold.  For the full mass range, the best-fit fold has a width $\sigma=0.033$ dex, $7.7\%$ scatter, and bias $\mu=-0.027$ dex.  In the mass range $14.0<\log(M_{500c})<14.5$, the best-fit fold has a width $\sigma=0.025$ dex, $5.7\%$ scatter, and bias $\mu=-0.021$ dex.

Putting this intrinsic scatter in context, luminosity ($L_X$)-based methods with excised cores typically have errors in the $15\%-18\%$ range \citep{2007ApJ...668..772M, 2018MNRAS.473.3072M}, while methods that utilize well-sampled clusters with high spatial and spectral resolution, such as a $Y_X$ approach, yield $5\%-7\%$ intrinsic scatter \citep{2006ApJ...650..128K}.  {A straightforward power law scaling relation relating mass to core-excised luminosity of the \tng cluster sample recovers the approximate expected scatter:   $14.6\%$ when the outer aperture has a modest $3\%$ error with $R_\mathrm{500}$, and $22.2\%$ when the outer aperture has a $5\%$ error with $R_\mathrm{500}$.  See Pop et al., \textit{in prep.}, for more information on the \tng cluster sample scaling relations.} 

Here, we have used low-resolution spatial information with no spectral data and {have achieved a $>20\%$ improvement over a global luminosity approach.  {In practice, a single, low-scatter model could be selected for an application of this method, implying that an improvement closer to $50\%$ is possible.}

As larger cosmological hydrodynamical simulations with more massive clusters become available, the scatter and bias of mass predictions may also be reduced by training on a cluster sample with a flat mass function \citep[as is used in][]{2015ApJ...803...50N} that more accurately describes the high mass cluster population. 

\begin{figure}[]
	\begin{center}
		\includegraphics[width=0.5\textwidth]{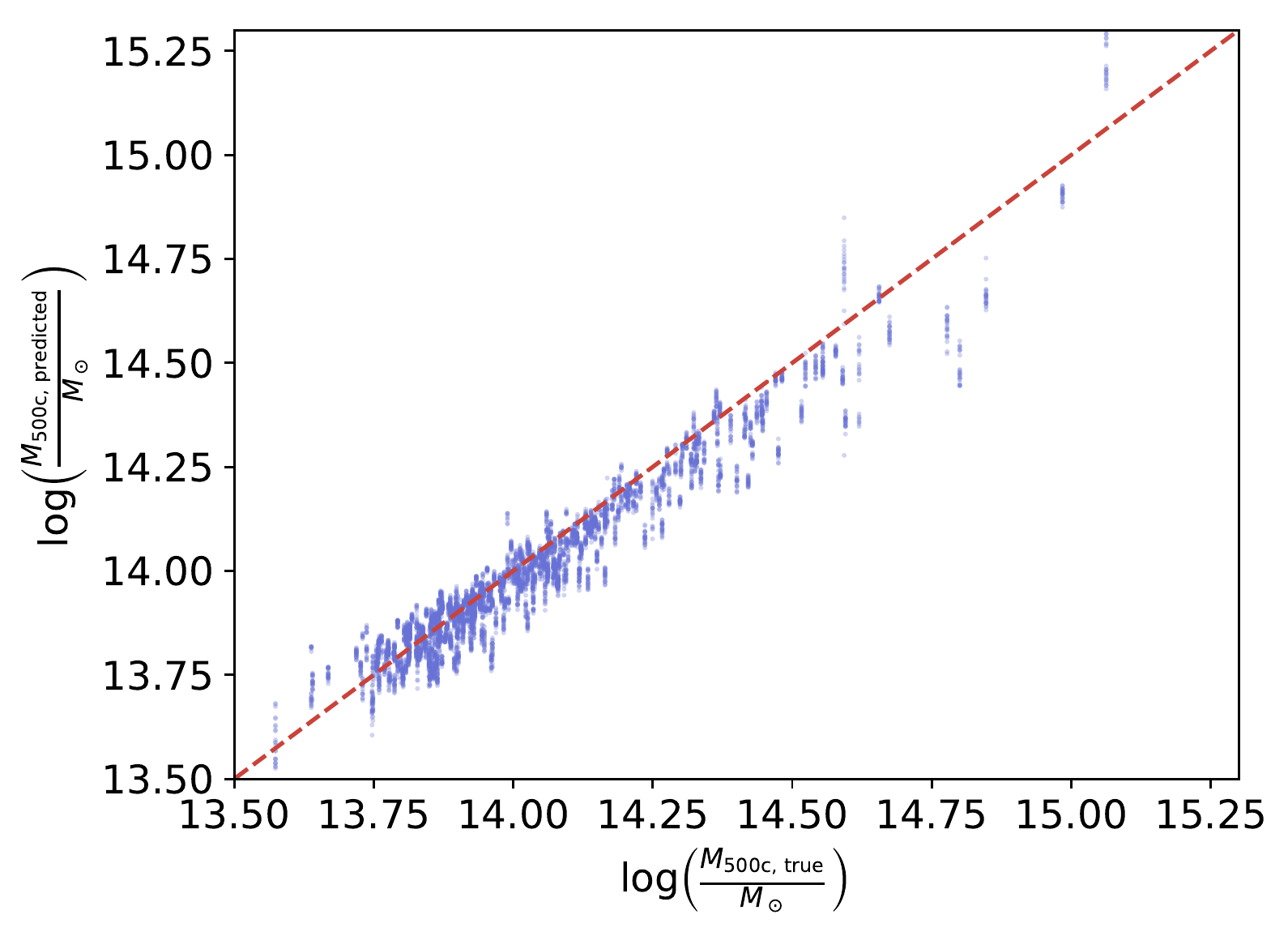} \\
	\caption[]{Predicted mass as a function of true mass.  The distribution has low intrinsic scatter (11.6\%) and a small negative bias ($-0.022$ dex).  The tendency to predict toward the mean  \textemdash{} {} overpredicting low mass clusters and underpredicting high mass clusters  \textemdash{} {} can be mitigated by carefully curating a training set that extends well beyond the mass range of test clusters. }
       	\label{fig:scatter}
      	\end{center}
\end{figure}
\begin{figure}[]
	\begin{center}
		\includegraphics[width=0.5\textwidth]{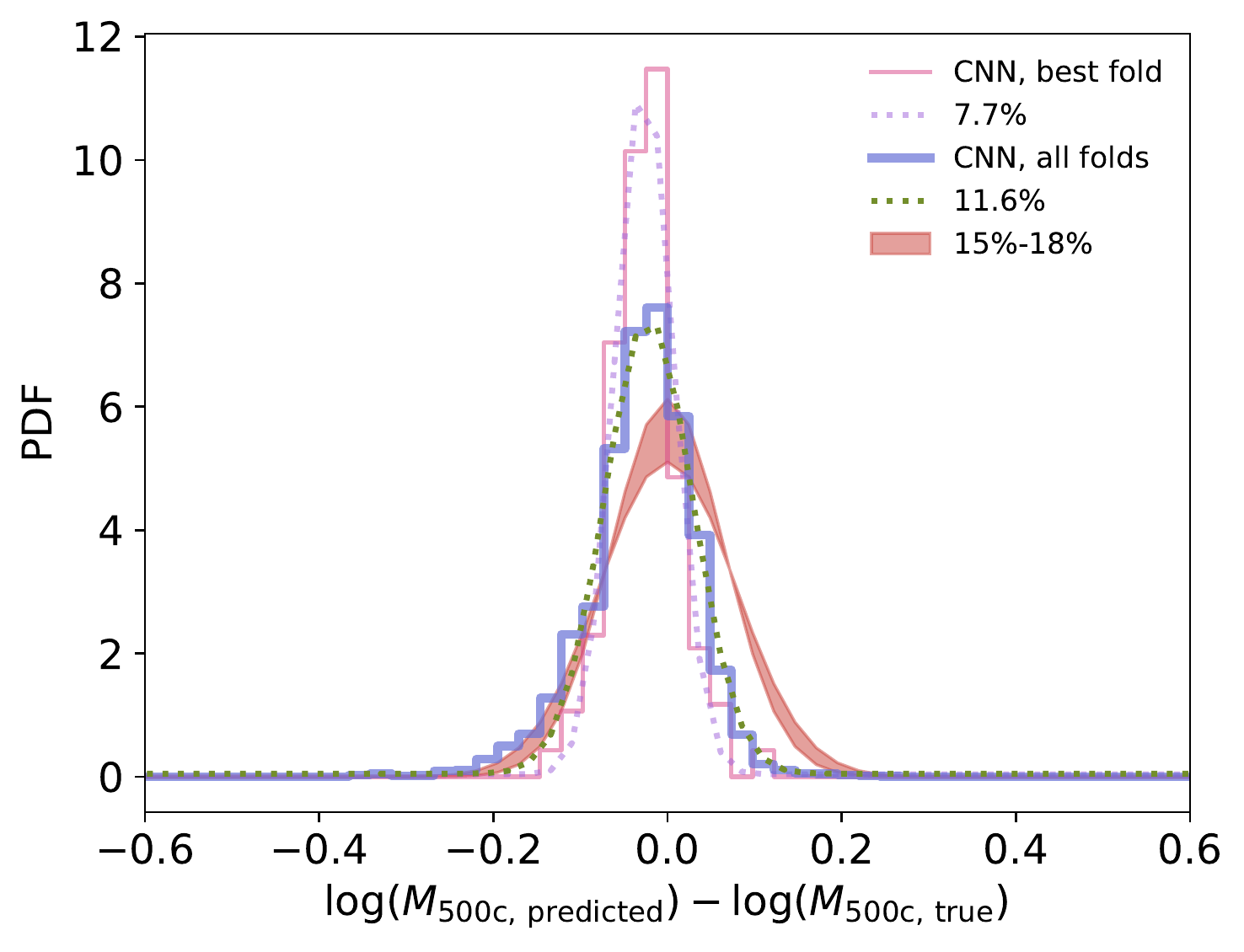} \\
	\caption[]{PDF of mass error (blue solid) given by $\log(M_\mathrm{predicted})-\log(M_\mathrm{true})$.  The full sample error distribution best fit Gaussian (green dash) has standard deviation $\sigma=0.051$ dex (11.6\% intrinsic scatter) and mean $\mu=-0.022$ dex.  
	The best-fit fold has a width $\sigma=0.033$ dex (7.7\% intrinsic scatter) and mean $\mu=-0.027$ dex (pink solid and purple dash).
	Core-excised $L_X$-based methods that use a single measure of cluster luminosity typically achieve a $15\%-18\%$ scatter (red band), while the $Y_X$ technique, which requires the full spatial and spectral observation, can yield a tighter $5\%-7\%$ scatter.
	Our CNN approach uses low-resolution spatial information to improve mass estimates over one based on a single summary parameter, $L_X$.}
       	\label{fig:errpdf}
      	\end{center}
\end{figure}

\newpage

\section{Interpreting the Model with DeepDream}
\label{sec:interpret}

\begin{figure}[]
	\begin{center}
		\includegraphics[width=0.45\textwidth]{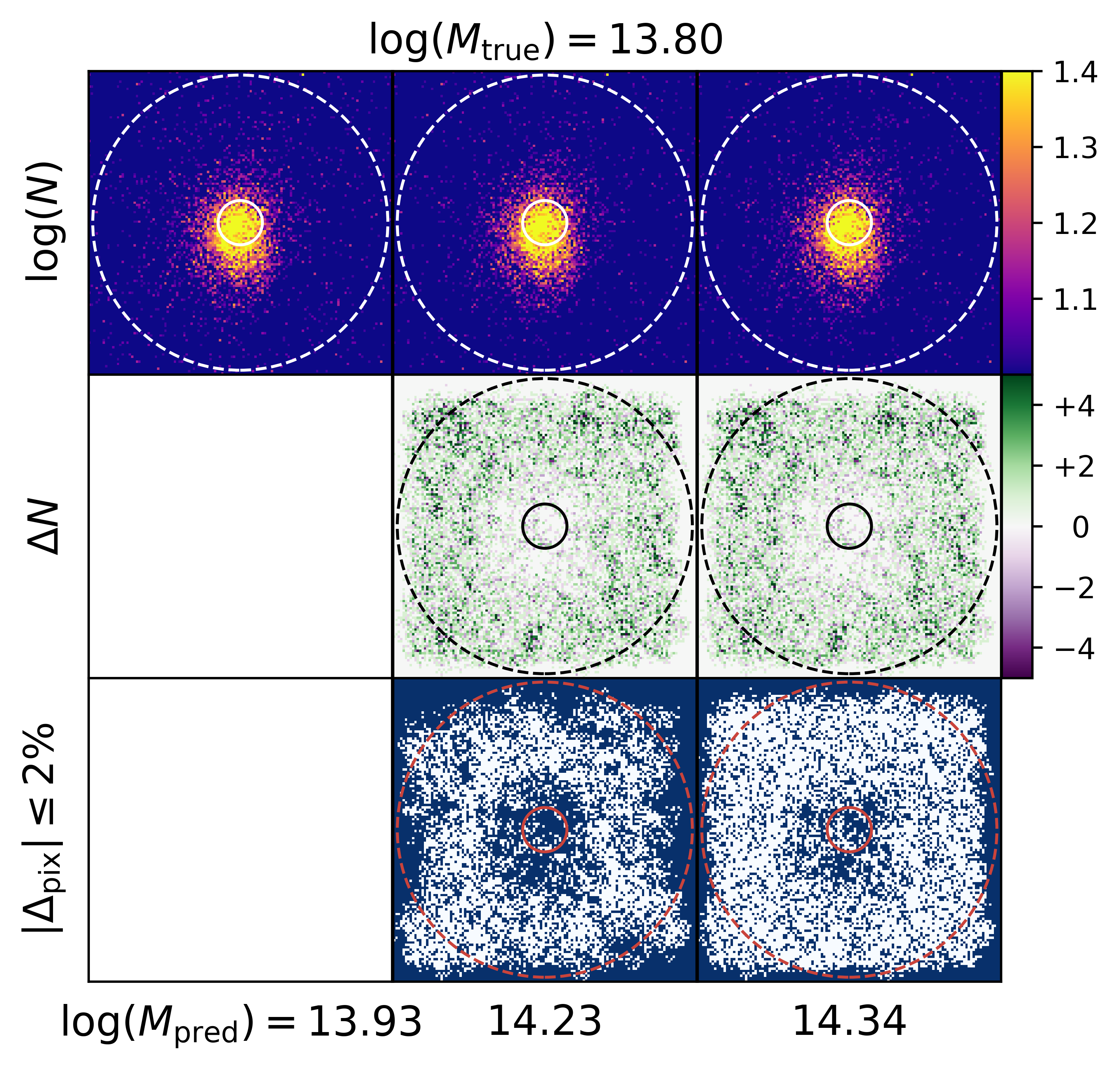} \\ %cluster 37
		\includegraphics[width=0.45\textwidth]{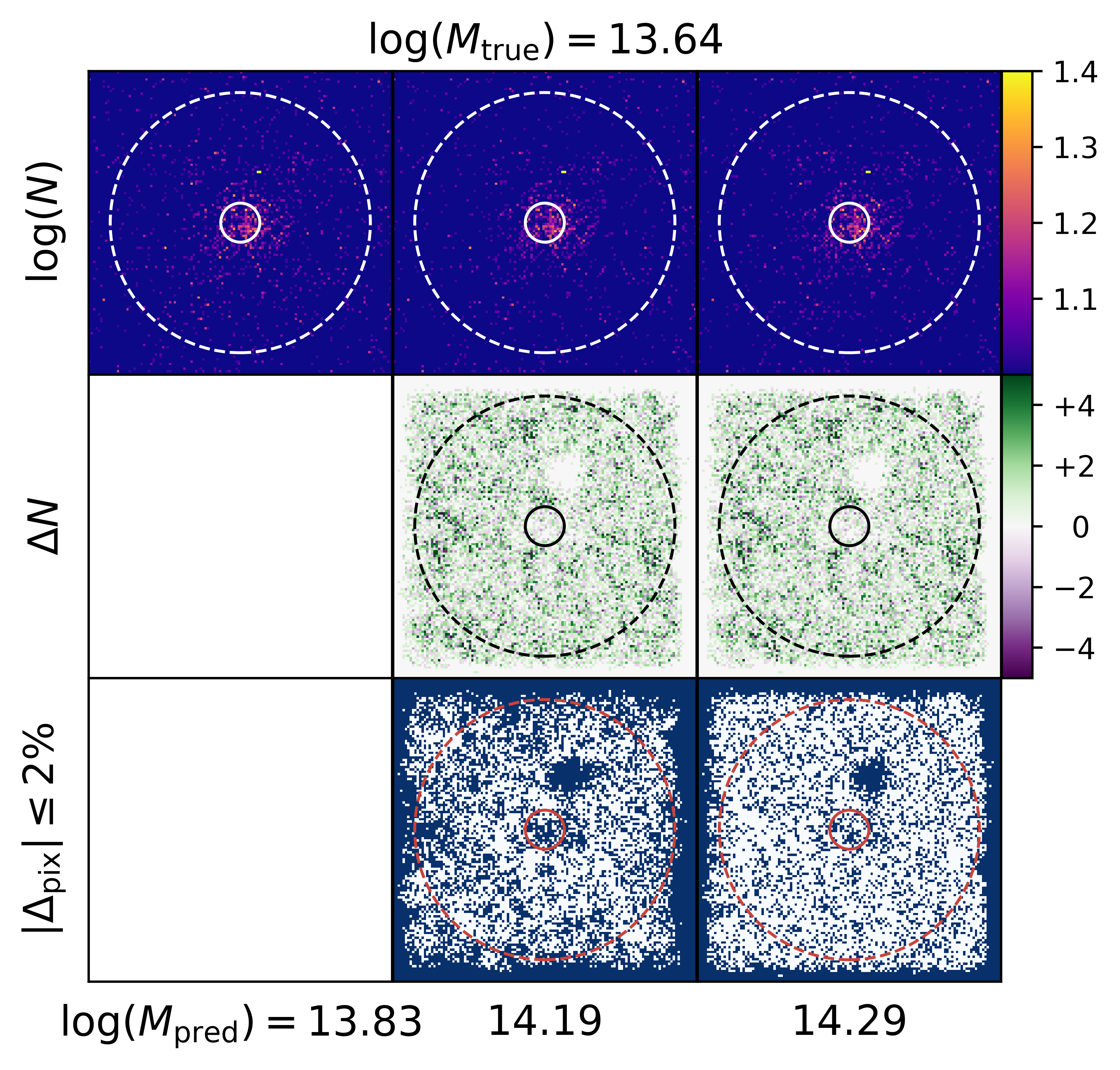} \\ %cluster 116
	\caption[]{Top panel:  a typical cluster's evolution over  two iterations in the DeepDream-inspired tool  for interpreting the CNN.
	Top row:  the original input image ($\log(N)$) of the cluster (left) is perturbed through two iterations (center and right) to increase the apparent cluster mass.
	Middle row:  changes in photon count ($\Delta N$) for each iteration (center and right) shows that the CNN tends to add photons roughly in a ring between $0.15 \, R_\mathrm{500c}$ (solid circle) and $1.0 \, R_\mathrm{500c}$ (dashed circle).   
	Bottom row:   dark pixels show the regions for which there is a small photon change ($\Delta N / N \leq 2\%$).
	Bottom panel:  Same as top panel, but for  one notable cluster for which the CNN misidentifies the cluster core above and to the right of the true cluster core.  This is highlighted in the bottom right image, where an off-center circular region has small photon count change. }
       	\label{fig:masschange}
      	\end{center}
\end{figure}

Convolutional Neural Networks are notoriously difficult to interpret.  To understand the method, we use {an approach inspired by DeepDream \citep[e.g.~][]{deepdream}.}  Google DeepDream uses gradient ascent applied to the input image pixels, asking the question ``What changes in the input image will result in a significant change in the classification of this image?''   Often, DeepDream is used to change the classification of a picture or photograph.  Our implementation differs in that our model regresses an output mass label, so we will be asking ``What changes in the input cluster image will result in a mass change of this image?''\footnote{For more details on visualizing filters of CNNs implemented in Keras, see  \cite{cnn_see}.}

We define our loss function  as the value of the final neuron output (the cluster mass) and compute the gradient of the input image with respect to this loss.  The gradient is a $128\times128$ single-color image and is calculated by finding the changes needed in each of the 16,384 pixels to maximize the cluster mass.  It is the change to the input image that  maximizes the loss function, in other words, adding the gradient to the input image results in a cluster that the trained CNN model interprets as being more massive.

{Adding the image gradient can introduce nonphysical properties to the image, including pixels with noninteger and negative photon counts.  To correct for this, we} impose physically-motivated constraints on the input image plus gradient:  pixels with negative photon counts are set to $0$ and pixels with non-integer values are rounded to the nearest integer.  It should be noted that these physically-motivated constraints do not significantly affect the CNN's prediction of the new cluster mass, nor do they significantly affect the plots or results presented here.

Figure \ref{fig:masschange} shows two sample clusters, including the input image, the gradients, and the updated cluster images for 2 iterations of this process.   The trained model typically adds photons outside of $\approx 0.2\,R_{500c}$ but ignores the core region of the cluster that is known to have large scatter with cluster mass \citep{2007ApJ...668..772M, 2018MNRAS.473.3072M}.

Figure \ref{fig:lumprofile} shows the fractional photon change, $\Delta N/N$, for a representative sample of clusters.  This is given by the ratio of photons in the iterated image (original image plus gradient) to the photons in the original image.  The 2-dimensional result is binned by radius.  The CNN ignores the central $\approx0.2\,R_{500c}$ of the cluster, typically adding photons outside of this region.  In the second iteration, photons are added even further from the cluster center.  This tool tends not to add photons near the edge of images, suggesting that some edge effects come into play.

\begin{figure}[]
	\begin{center}
		\includegraphics[width=0.5\textwidth]{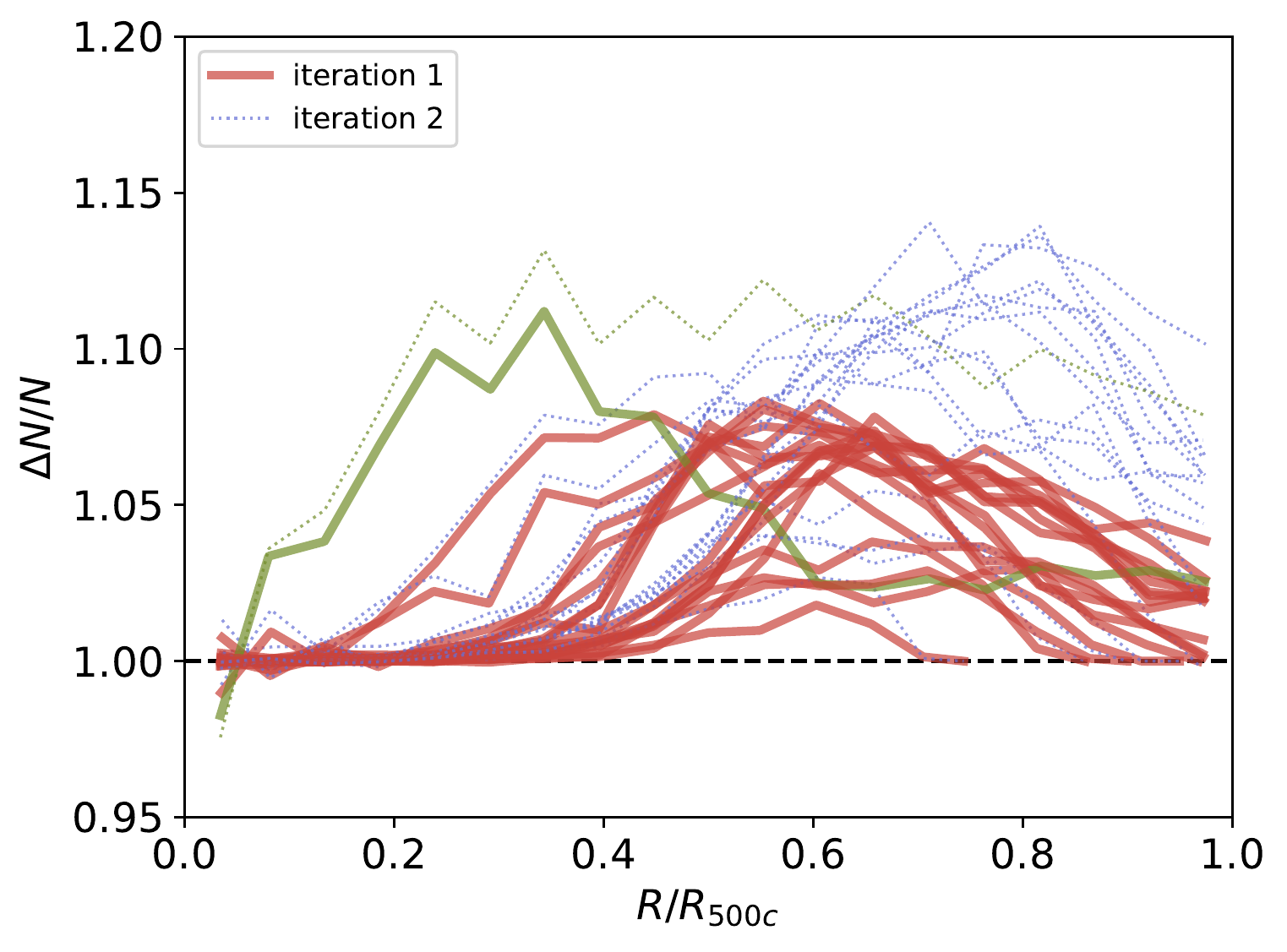} \\
	\caption[]{The fractional change in photons ($\Delta N/N$) as a function of projected distance from the cluster center ($R/R_{500c}$) for a representative sample of clusters for the DeepDream interpretation of the trained CNN.  The first iteration  (solid red) adds photons beyond $\approx 0.2 \, R_{500c}$, while the second iteration (blue dotted) increases the photon count at larger radii.  This suggests that the CNN has learned to excise cores, which have been shown to have large scatter with $M_{500c}$.  One notable exception (green solid and dotted) is the cluster highlighted in the bottom panel of Figure \ref{fig:masschange}.  The gradients for this cluster have an empty region near the cluster's off-center bright region; this cluster is discussed in more detail in Section \ref{sec:interpret}.  Interpretation tools such as the one presented here can be used to understand the features used by a CNN to regress cluster mass.}
       	\label{fig:lumprofile}
      	\end{center}
\end{figure}

One notable exception is shown in the bottom panel of Figure \ref{fig:masschange} and highlighted in Figure \ref{fig:lumprofile}.  In this case, the CNN adds photons near the core region.  Further inspection of this cluster reveals two bright, off-center pixels above and to the right of the cluster center.  The CNN incorrectly interprets this bright region as the cluster core, perturbing the image by adding photons surrounding it.  {These bright photons likely originated from nonphysical, high-density, ionized cold particles which are not easily filtered out.  In analysis of real \chandra observations, point sources like these would be removed in preprocessing.  } When applying the CNN mass estimate method to a sample of observed clusters, our interpretation may be useful in {identifying, categorizing, and understanding} outliers such as this cluster.

{{Our data-driven model, calibrated on a simulation, has achieved small bias and scatter without invoking assumptions about hydrostatic equilibrium or hydrostatic mass bias.  The model has learned to excise cores, ignoring the region of the cluster most directly affected by feedback physics, which is difficult to model.  Although the global intercluster medium  profiles are relatively unaffected by still poorly-understood cluster core physics \citep{2016MNRAS.459.2973S},  \tng-calibrated mass estimates will depend on the physics of cluster outskirts (see \cite{2018arXiv181000890W} for a recent review).  {Encouragingly,  \tng clusters recover the expected X-ray scaling relations  (Pop et al., \textit{in prep.}) and metallicity profiles \citep{2018MNRAS.474.2073V}, suggesting that the  outskirts of these simulated clusters are in good agreement with observed clusters.} To confirm that no significant bias is introduced by the gas physics modeling, weak lensing mass estimates of a well-studied cluster sample can provide an important cross check.}}
 
\section{Conclusion}
\label{sec:conclusion}
 
We have presented a method for inferring cluster masses from \chandra mock observations of galaxy clusters.  The mock observations are built from 329 massive clusters within a mass range of $M_{500c}=10^{13.57}$ to $M_{500c}=10^{15.06}$ from the \textit{TNG300} simulation.  The mass proxy uses a Convolutional Neural Network (CNN) with three pairs of convolutional and pooling layers followed by three fully connected layers.  The model is trained to learn cluster mass from a low spatial resolution, single-color X-ray image.  Our approach shows that a low resolution X-ray image of a galaxy cluster can be used to predict the mass with low scatter (12\%)  and low bias (-0.02 dex) without invoking assumptions about the hydrostatic mass bias.  

The scatter may improve as larger simulations become available, providing catalogs that better sample the high mass region of the halo mass function.  Ultimately, this method may be trained on simulations to predict the masses of \chandra-observed clusters.

{Machine learning tools are commonly viewed as black boxes that produce an answer without an interpretation, and it can be particularly difficult to glean a physical understanding from deep learning methods.  We aim to remedy this by studying our trained model with an approach inspired by Google DeepDream}.  We calculate a gradient necessary to perturb an input cluster image so that the trained CNN will increase the mass estimate.  {We find that the trained CNN is most sensitive to photons from the cluster outskirts and ignores the inner ($R \lesssim 0.2 R_{500c}$) regions of the cluster, in agreement with what has been found by more conventional statistical analyses of galaxy clusters.}  The method can be useful in providing a physical {interpretation of the features of the cluster sample that are relevant for predicting masses from X-ray images.}

{As new, large cluster samples become available,  new data-driven methods} will need to be developed to take advantage of these data sets.  For example, the upcoming eROSITA mission \citep{2012arXiv1209.3114M} is estimated to find $\approx 93,000$ galaxy clusters with masses larger than $10^{13.7}\Msolarh$ \citep{2012MNRAS.422...44P, 2018MNRAS.481..613P}. 
New tools, such as the CNN-based mass proxy presented here, can be useful  in analyzing and understanding large observational data sets.  Utilizing CNNs to infer X-ray masses of the eROSITA cluster sample, however, will require a much larger training sample of simulated images spanning a wide dynamic range in cluster mass.  As bigger simulated cluster catalogs become available, CNNs may prove to be a powerful tool for analyzing and understanding large cluster observations.

\acknowledgments{ 
We thank Dominique Eckert, Sheridan Green,  Francois Lanusse, Paul La Plante, Junier Oliva, and Kun-Hsing Yu for their helpful feedback on this project.
 }\\ \\

\bibliography{../../references}
\bibliographystyle{apj}

\end{document}